\documentclass[twocolumn,preprintnumbers,showpacs,amsmath,amssymb]{revtex4}
\usepackage{amssymb}
\usepackage{calrsfs}

\begin{document}
\title{Comment on \textquotedblleft A short impossibility proof of quantum
bit commitment\textquotedblright }
\author{Guang Ping He}
\email{hegp@mail.sysu.edu.cn}
\affiliation{School of Physics and
Engineering, Sun Yat-sen University, Guangzhou 510275, China}

\begin{abstract}
In a recent letter (Phys. Lett. A 377 (2013) 1076), the authors presented
an impossibility proof of quantum bit commitment, which attempted to cover all
possible protocols that involve both quantum and classical information. Here
we show that there are many errors in the proof, thus it fails to exhaust
all conceivable protocols.
\end{abstract}



\maketitle

\newpage


While it is widely believed that unconditionally secure quantum bit
commitment (QBC) is impossible, the completeness of previous impossibility
proofs have been continually challenged. As pointed out in a recent letter
\cite{qi715}, \textquotedblleft the debate can be only settled with an
appropriate formulation of the problem, which is sufficiently powerful to
include all possible protocols in a single simple mathematical
object\textquotedblright . Ref. \cite{qi715} made a good attempt towards
this direction. Unfortunately, we found that the goal is still not
accomplished, as their proof contains the following problems.

\section{Wrong modelling of the\ history of classical information}

At the beginning of Section 4.1 of Ref. \cite{qi715}, the authors wrote that
\textquotedblleft $s_{l}=i_{0}i_{1}...i_{l}$ will represent the history of
classical information with $i_{2k-1}$ denoting the outcome of Bob's quantum
operation at step $k$ (which is the same as Alice's classical input at
Alice's step $k$) and $i_{2k-2}$ for $k>1$ represents Bob's input classical
information (which is Alice's output at step $k-1$)\textquotedblright . That
is, they think that Bob's (Alice's) output always equals to Alice's (Bob's)
input. This point may look fine at the first glance, because in the 1st
paragraph of Section 3.7 the authors defined $s_{l}$\ (and therefore all $i$%
's included in $s_{l}$) as the classical information being openly exchanged.
However, we should note that there can be protocols in which some outcomes
are allowed to be kept secret without being exchanged. Meanwhile, these
secret information can also affect the participants' choices of quantum
operations. For example, at a certain step $k_{0}$\ Bob's outcome can
include both $i_{2k_{0}-1}$ and $i_{2k_{0}-1}^{\ast }$, where $i_{2k_{0}-1}$
is publicly exchanged and becomes Alice's input while $i_{2k_{0}-1}^{\ast }$
is kept secret to himself. At a later step $k_{1}$, Bob can choose which
quantum operation to apply, depending not only on the classical information
already exchanged, but also on $i_{2k_{0}-1}^{\ast }$ (e.g., on a certain
comparison result between $i_{2k_{0}-1}^{\ast }$ and Alice's recent output). Thus $%
i_{2k_{0}-1}^{\ast }$ becomes a part of Bob's own input at this step. In
such protocols, \textit{one participant's output is not always the same as
the other's input at each step}, as opposed to the case the authors studied.

As a consequence, in the general case the\ history of classical information $%
s$ should include not only all $i$'s being exchanged, but also all $i^{\ast
} $'s being kept secret. Such an $s$ is not always known completely to
either Alice or Bob. Each of them can only know a different part of $s$. A
cheating strategy must conform to all $i$'s, i.e., the communication
interface, as mentioned at the end of the 1st paragraph of their Section 4.
But this is not enough. A successful cheating should also agree with $%
i^{\ast }$'s of the other participant. Therefore, it becomes doubtful
whether a dishonest Alice has sufficient information to calculate the
unitary transformation $\mathcal{P}$\ in Eq. (69), as the related equations
in their Section 5.1 all depend on the history of classical information $s$.
(Note that the authors dropped the index $s$ for simplicity, as they
mentioned in the paragraph below Eq. (68).) Without $\mathcal{P}$, Alice's
cheating procedure described below Eq. (72) cannot be performed.

Some might argue that in a protocol all participants can measure only the
information that are required to be publicly announced. All other secret can
be kept on the quantum level without collapsing into classical information,
so that $i^{\ast }$'s can be treated as a part of the quantum system being
exchanged and does not need to be included in $s$. However, while
classical information is a special case of quantum ones, it is so special
that it can be cloned perfectly, cannot be altered, and there is only one
measurement basis for reading the information, no other nonorthogonal bases
exist. On the contrary, quantum information does not have these
particularities. Thus classical information cannot always be replaced with
quantum ones, especially when there can be dishonest participants who want
to alter the content or perform measurements in other bases. Delaying the
measurement on one's own quantum system may not always be a benefit for an
honest participant (at least, even though it is indeed a benefit in some
particular protocols previously proposed, there still lacks of such a
general proof in literature, including Ref. \cite{qi715}). Especially, as
described in the 1st paragraph of Section 3.7, the quantum system is
exchanged back and forth between the participants. Therefore in a
well-designed protocol, an honest participant's measuring his own quantum
system can help him gain more information from the system, and keeping
these information secret will put more limits on the other participant's
freedom on altering the system dishonestly in later steps. (Note that it is
\textit{not} the purpose of our current comment to discuss whether the
\textit{conclusion} \textquotedblleft unconditionally secure QBC is
impossible\textquotedblright\ is correct. Instead, it is to discuss whether
the current \textit{proof} is sufficient to lead to this conclusion. For
this reason, here we will not go into detail on how to construct such a
protocol, and would rather leave it elsewhere.) On the other hand, if an honest
participant announces openly all the classical information that he obtained
from his measurements, then he is surely easier to be cheated since he lets his
opponent know too much. Therefore, if $s$ is taken merely as the classical
information being exchanged, while the secret information $i^{\ast }$'s are
not included, then not all the measurements in the protocols are fully
considered, so that the corresponding impossibility proof cannot be regarded
as general.

Also, it might be the intention of the authors that these secret information
could be treated as the content of the quantum memories, so that they do not
need to be included in the classical information $s$. But we must note that
the quantum memory was merely mentioned briefly in Sections 3.2 and 3.3 of
Ref. \cite{qi715}. After that, the authors seemed to forget its role. The
contribution of its content on the form of the cheating operations was never
formulated in a rigorous mathematical manner. Especially, the change of the
memory content was not considered.

\section{Loophole on the correspondence between strategies and conditioned
combs}

After the proof of Theorem 2 in Section 3.7, it was claimed that as a
consequence of the theorem, \textquotedblleft single-party strategies in a
protocol are in one-to-one correspondence with conditioned
combs\textquotedblright . However, let us look into the details of the proof
of Theorem 2. It said that \textquotedblleft \textit{if} we apply the von
Neumann-L\"{u}ders measurements $\{I_{S_{2k}}\otimes \left\vert
s_{2k}\right\rangle \left\langle s_{2k}\right\vert \}$\ on the input space $%
\mathcal{H}_{2k}$\ before channel $\mathcal{C}_{k}$, followed by $%
\{I_{S_{2k+1}}\otimes \left\vert s_{2k+1}\right\rangle \left\langle
s_{2k+1}\right\vert \}$\ on the output space $\mathcal{H}_{2k+1}$\ after
channel $\mathcal{C}_{k}$, we obtain the conditioned quantum operations $\{%
\mathcal{C}_{i_{2k+1}}^{s_{2k}}\}$\textquotedblright . That is, the
correspondence between a strategy and a conditioned comb is valid only
\textit{if} the mentioned measurements are performed. But what if the
measurements are applied in other bases, or not performed at all? Then there
will be no guarantee that the correct conditioned quantum operations will be
obtained.

Note that \textquotedblleft whether QBC can be secure\textquotedblright\ and
\textquotedblleft whether the current proof is sufficiently
general\textquotedblright\ are two different stories. That is, on one hand,
it is true that a protocol is insecure if it can be proven that a successful
cheating strategy still exists even when the cheater honestly performs the
measurement required in this particular step. While on the other hand, the
model of QBC cannot be considered general if there is no proof showing that
the cheater always has to perform this measurement honestly. In the
presented proof, as shown in their Fig. 2, the measurement of the quantum
tester is performed at the opening stage only, instead of being performed
after each step $k$ ($k=0,...,N-1$). Therefore, while the protocol requires
the strings $s_{2k}$, $s_{2k+1}$\ to be announced classically, there is
nothing to ensure that both participants indeed obtain all these strings by
the honest measurements $\{I_{S_{2k}}\otimes \left\vert s_{2k}\right\rangle
\left\langle s_{2k}\right\vert \}$\ and $\{I_{S_{2k+1}}\otimes \left\vert
s_{2k+1}\right\rangle \left\langle s_{2k+1}\right\vert \}$. It is possible
that they merely announce $s_{2k}$, $s_{2k+1}$\ by guess without
measurements, or by using other dishonest measurements.

For this reason, the proof that the authors presented to Theorem 2 is valid
only when all internal participants always perform the measurements
honestly. Thus it works fine if we apply Eq. (54) of Ref. \cite{qi715}
merely to formulate quantum cryptographic protocols in which all internal
legitimate participants are willing to collaborate honestly against external
cheating. The well-known quantum key distribution is a typical example of
such protocols. On the other hand, in cryptographic tasks where some of the
internal legitimate participants may attempt to cheat (e.g., QBC), the
measurements required in the proof of Theorem 2 may not be performed
honestly. In this case, the correspondence between strategies and
conditioned combs does not necessarily holds, so that Eq. (54) is not
sufficiently general to represent all possible protocols \textit{where there
are dishonest internal participants}.



\section{Who should perform the last move}

Also in the 1st paragraph of Section 4.1, the authors claimed that
\textquotedblleft at the end of the commitment stage we can assume without
loss of generality that Alice performs the last move\textquotedblright\
because \textquotedblleft for a protocol where the last move is Bob's, we
can always add a null move\textquotedblright . This is a tricky part of
their impossibility proof, which makes it easier to come up with a cheating
strategy for Alice. All these who had attempted to construct unconditionally
secure QBC would know that one of the biggest difficulty is to force
measurements, especially, to force a dishonest Alice to finish her
measurements during the commitment stage. It is surely favorable to Alice if
she is assumed to perform the last move and she does not use a null one, as
Bob can no longer check whether Alice has performed this step honestly (or
whether it is indeed a null move) before the end of the commitment stage.
For example, if an honest Alice is supposed to finish some measurements in
the last move, a dishonest Alice will then be able to delay the measurements
until the beginning of the opening stage, because there is no Bob's move in
between to check Alice's behavior. Thus we can see that \textit{assuming
Alice to perform the last move is not appropriate}.

\section{Misuse of the concealment condition}

In the last paragraph of the left column of page 9, the authors wrote that
\textquotedblleft a protocol that is not concealing at step $k$ is also not
concealing at any following step\textquotedblright . This statement could be
wrong, depending on how \textquotedblleft not concealing\textquotedblright\
is defined here. It may be true if \textquotedblleft not
concealing\textquotedblright\ means that $\rho ^{(0)}$\ and $\rho ^{(1)}$
(the density matrices\ of the states at Bob's side at this step
corresponding to the committed values $b=0$ and $b=1$, respectively) are
rigorously orthogonal to each other. But it is not true if it\ means that $%
\rho ^{(0)}$\ and $\rho ^{(1)}$ do not satisfy the concealment condition Eq.
(60) rigorously. In fact, \textit{Eq. (60) does not have to be satisfied at
any step of the protocol.} It is necessary only for the last step before the
opening. This is because in the middle of the commitment stage of a
well-designed protocol, Bob will still be
subjected to other security checks at later steps, where he may be required
to measure his quantum system in other bases which are different than the
one that could distinguish $\rho ^{(0)}$\ and $\rho ^{(1)}$. If the protocol
that is partially not concealing at step $k$, i.e., Eq. (60) is not
satisfied while $\rho ^{(0)}$\ and $\rho ^{(1)}$ are not rigorously
orthogonal either, then Bob's measurement for distinguishing $\rho ^{(0)}$\
and $\rho ^{(1)}$ could cause a detectable disturbance to the quantum
system, so that he will be caught in the security checks at later steps.
With this method, while the protocol is not concealing at a middle step $k$, Bob is
held back from applying the necessary measurement to distinguish $\rho
^{(0)} $\ and $\rho ^{(1)}$.
Thus it is insufficient for the impossibility proof when the authors
only \textquotedblleft focus on the last step $N$ before the
opening\textquotedblright . Consequently, their results thereafter become
unreliable.

Similarly, the 1st sentence of the last paragraph of their Section 5.2 said
that \textquotedblleft for a protocol with unbounded number of rounds, the
conditions of $\varepsilon $-concealment and $\delta $-closeness are still
given by Eqs. (60) and (65), respectively\textquotedblright . We can see
that the authors also missed to notice the possibility that a secure
protocol does not have to be concealing at the middle stage, as long as
there is proper security checks in later steps that can prevent Bob from
distinguishing $\rho ^{(0)}$\ and $\rho ^{(1)}$ at this stage. Therefore,
their proof is insufficient to cover all protocols with
unbounded number of rounds either.

\section{Other errors}

There are also other problems in Ref. \cite{qi715} which seem relatively
less critical but still should not be ignored, as enlisted below.

\subsection{Probability of failure}

At the end of Section 2.1, the authors said that Bob's measurement will
result \textquotedblleft in a failure, e.g. due to the detection of an
attempted cheat. Again, in a well-designed protocol the probability of
failure should be vanishingly small\textquotedblright . This is bewildering.
When Alice cheats, the probability of failure in a well-designed protocol
should \textit{not} be small, otherwise the protocol is surely insecure.

\subsection{Tester for an honest Bob}

In the 1st paragraph of Section 3.4, it stated that \textquotedblleft a
dishonest Bob will perform a tester to distinguish Alice's strategies before
the opening\textquotedblright . This statement could sound misleading, as it
seems to suggest that an honest Bob will not perform the tester. But if this
is true, then the QBC being studied are merely those protocols in which the
measurements of an honest Bob are excluded. That is, a large class of QBC
protocols with measurements will be left out from their proof.

\subsection{Concealment versus abortion}

In the 2nd paragraph of its Summary, the authors claimed that
\textquotedblleft concealment is defined regardless abortion, namely Bob
cannot detect the bit value anyway, whether Alice catches him or
not\textquotedblright . This condition is too strong. If Bob can
learn the bit but Alice can catch him whenever he manages to do so, then the
protocol is generally still regarded as secure. \\

In summary, there are many logical problems in Ref. \cite{qi715}. Thus it
becomes doubtful whether the existing impossibility proof of unconditionally
secure QBC is sufficiently general to cover all possible protocols. Still,
comparing with a lot of previous impossibility proofs, the current one is more
valuable as it attempts to construct the proof with an unambiguous detailed
presentation, so that the (im)possibility of QBC can be discussed in a precise way. \\


The work was supported in part by the NSF of China under grant No. 10975198,
the NSF of Guangdong province, and the Foundation of Zhongshan University
Advanced Research Center.

\end{document}